\documentclass[aps,prl,twocolumn,superscriptaddress]{revtex4}
\usepackage{epsf,graphicx}
\usepackage{amssymb}
\usepackage{amsmath}
\usepackage{latexsym,bm,array,amsfonts,multirow}
\usepackage{color}
\usepackage{ulem}
\begin{document}

\title{Pressure-induced self-doping and Fermi surface reconstruction in UAs$_2$}
\author{Zhenchao Wu}
\affiliation{Beijing National Laboratory for Condensed Matter Physics and
	Institute of Physics, Chinese Academy of Sciences, Beijing 100190, China}
\affiliation{School of Physical Science and Technology $\&$ Key Laboratory of Quantum Theory and Applications of MoE, Lanzhou University, Lanzhou 730000, China}
\author{Yingying Cao}
\email[]{caoyingying@hnas.ac.cn}
\affiliation{Institute of Quantum Materials and Physics, Henan Academy of Sciences, Zhengzhou 450046, China}
\author{Yi-feng Yang}
\email[]{yifeng@iphy.ac.cn}
\affiliation{Beijing National Laboratory for Condensed Matter Physics and Institute of
	Physics, Chinese Academy of Sciences, Beijing 100190, China}
\affiliation{University of Chinese Academy of Sciences, Beijing 100049, China}
\date{\today}

\begin{abstract}
Superconductivity has recently been reported in the heavy-fermion compound $\mathrm{UAs_2}$ under pressure, with the highest $T_c$ among uranium-based correlated $5f$-electron superconductors. To elucidate its microscopic origin, we investigate its electronic structure using density functional theory combined with dynamical mean-field theory (DFT+DMFT). At ambient pressure, our calculations reproduce the characteristic Kondo-lattice electronic structure, with flat hybridization bands near the Fermi energy around the $\Gamma$ and M points, in good agreement with angle-resolved photoemission spectroscopy (ARPES). Under pressure, we find a systematic transfer of electrons from the more localized $5f_{5/2}$ orbitals to the more itinerant $5f_{7/2}$ orbitals, while the total U-$5f$ occupancy remains nearly unchanged. This orbital-selective charge redistribution constitutes a pressure-induced self-doping effect that drives the $5f_{5/2}$ electrons from a localized Kondo regime toward a mixed-valence regime with enhanced charge fluctuations, leading to a dramatic reconstruction of the low-energy electronic structure. Remarkably, superconductivity emerges in the pressure range where the Fermi surface consists of two disconnected sheets with enhanced nesting, but disappears when they bend and merge into a corrugated three-dimensional cylinder. Our results provide an electronic-structure basis for understanding superconductivity in $\mathrm{UAs_2}$ and suggest that Fermi-surface nesting and charge fluctuations may contribute to the enhanced superconducting $T_c$, pointing to a possible distinction from conventional heavy-fermion superconductors.
\end{abstract}

\maketitle

\textit{Introduction.---}Unconventional superconductivity in $f$-electron systems has long been associated with the intricate interplay between magnetism, electronic correlations, and Fermi surface reconstruction \cite{StewartRMP1984,GegenwartNP2008,MathurNat1998,PaschenNat2004}. Uranium-based compounds are particularly intriguing in this context due to the dual nature of their $5f$ electrons, which lie at the boundary of localization and itinerancy \cite{SaxenaNat2000,JoyntRMP2002}. This duality gives rise to a wide range of emergent phenomena, including heavy-fermion behavior, magnetically mediated pairing, and possible spin-triplet superconductivity, as exemplified by $\mathrm{UTe_2}$ \cite{RanSci2019,AokiJpsj2019,XuPRL2019}. Recent high-pressure studies on uranium-based compounds have further expanded this landscape, with $\mathrm{UAu_2}$ emerging as a candidate system exhibiting superconductivity and possible multicomponent odd-parity pairing under pressure \cite{O’NeillPnas2022}. Beyond these recently discovered superconducting candidates, uranium dipnictides $\mathrm{U}X_2$ ($X$ = P, As, Sb, Bi) have been extensively investigated over decades due to their intriguing magnetic and electronic properties \cite{AmorettiJmmm1984,Gerward1990,ChenPRL2019,GiannakisSciAdv2019}. These compounds exhibit antiferromagnetic order with relatively high Néel temperatures and pronounced electronic correlations \cite{FengPRB2021,MiaoNc2019,SiddiqueeNc2023}. Among them, $\mathrm{UAs_2}$ stands out as a correlated antiferromagnetic metal with clear signatures of $5f$–conduction electron hybridization and quasi-two-dimensional electronic structure \cite{laiprb2022,JiPRB2024}, making it a promising platform to explore correlation-driven superconductivity.

Recent high-pressure experiments have revealed a sequence of electronic and structural phase transitions in $\mathrm{UAs_2}$ under compression, as summarized in Fig.~1 \cite{Wen2026}. At ambient pressure, $\mathrm{UAs_2}$ exhibits antiferromagnetic order below $T_{\mathrm{N}} \approx 274$ K. With increasing pressure, the Néel temperature decreases monotonically to below 100 K and then abruptly collapses at a critical pressure of $P_{\mathrm{c}} \approx 20$ GPa. A pressure-induced structural transition then occurs, accompanied by the evolution from the tetragonal to orthorhombic crystal structure. Upon further compression, superconductivity emerges at low temperatures above $\sim$22 GPa and forms a dome, with a maximum onset $T_{\mathrm{c}} \approx 4.05$ K at around 26.8 GPa, the highest among uranium-based correlated 5$f$-electron superconductors. The superconductivity remains robust against external magnetic fields, with an upper critical field $\mu_{0}H_{\mathrm{c2}}(0) \sim 12$ T, exceeding the Pauli paramagnetic limit. In the normal state above the superconducting $T_{\mathrm{c}}$, non-Fermi-liquid (NFL) behaviors characterized by a linear-in-temperature resistivity were reported, indicative of a strange-metal regime. At higher pressures beyond 31.9 GPa, the superconductivity disappears and the system gradually recovers a Fermi-liquid (FL) ground state. Despite these experimental advances, the microscopic mechanism driving this pressure-induced superconductivity remains elusive. In particular, it remains unclear whether the superconducting phase can be understood within a conventional heavy-fermion framework and what gives rise to its relatively high $T_c$.

In this work, we employ density functional theory (DFT) \cite{2014WIEN2k,WIEN2k} combined with dynamical mean-field theory (DMFT) \cite{Georges1996RMP, Anisimov1997JPCM, Lichtenstein1998PRB, Kotliar2006RMP, Held2008JPCM, Haule2010PRB} to investigate the electronic structure of $\mathrm{UAs_2}$ and its evolution under pressure. At ambient pressure, the calculated spectral functions are consistent with available angle-resolved photoemission spectroscopy (ARPES) measurements, reproducing the characteristic flat bands near the Fermi level around the $\Gamma$ and M points that originate from the strongly correlated U-$5f$ electrons. Upon increasing pressure, we observe a pronounced redistribution of the U-$5f$ electron occupancy, with electrons transferring from the $j=5/2$ to the $j=7/2$ manifolds. This pressure-induced self-doping enhances charge fluctuations and drives a dramatic reconstruction of the Fermi-surface topology. Superconductivity emerges in the pressure range where the Fermi surface consists of two disconnected sheets with enhanced nesting. Our results reveal a close connection between pressure-induced self-doping, Fermi-surface topology, and superconductivity in UAs$_2$, and suggest that enhanced Fermi-surface nesting and charge fluctuations may provide a favorable electronic environment for unconventional superconductivity with elevated $T_c$.

\textit{Method.---}DFT calculations were performed using the full-potential augmented plane-wave plus local orbital method as implemented in the WIEN2K package with the Perdew–Burke–Ernzerhof exchange-correlation functional \cite{2014WIEN2k, WIEN2k, Perdew1996PRL}. The lattice parameters were taken from experiment \cite{Wen2026, Gerward1990}. Given the strong relativistic effects in 5$f$ electron systems, spin–orbit coupling (SOC) was explicitly included in the calculations. For simplicity, we only focus on the nonmagnetic state to see if useful informations may be extracted for understanding the non-Fermi liquid and the superconductivity.

DFT+DMFT approach was employed to treat the electronic correlations of U-5$f$ orbitals, using the hybridization expansion continuous-time quantum Monte Carlo approach as the  impurity solver \cite{Haule2007PRB}. To construct the DMFT Hamiltonian, we employed the Kohn-Sham bands within the energy window from -10 to 10 eV with respect to the Fermi energy. The local Coulomb interactions in U-5$f$ shell were described within the density–density approximation, with the on-site Hubbard interaction $U = 8$ eV and the Hund's rule coupling $J = 0.6$ eV following previous calculations for uranium oxides \cite{ShimEl2009,YinPRB2011,XuPRL2019}. The double counting is treated using the exact scheme \cite{Haule2010PRB,Haule2015PRL} and the spectral function was obtained using the maximum entropy method for analytic continuation of the self-energy \cite{Jarrell1996PR}.

\begin{figure}[t]
	\begin{center}
		\includegraphics[width=0.48\textwidth]{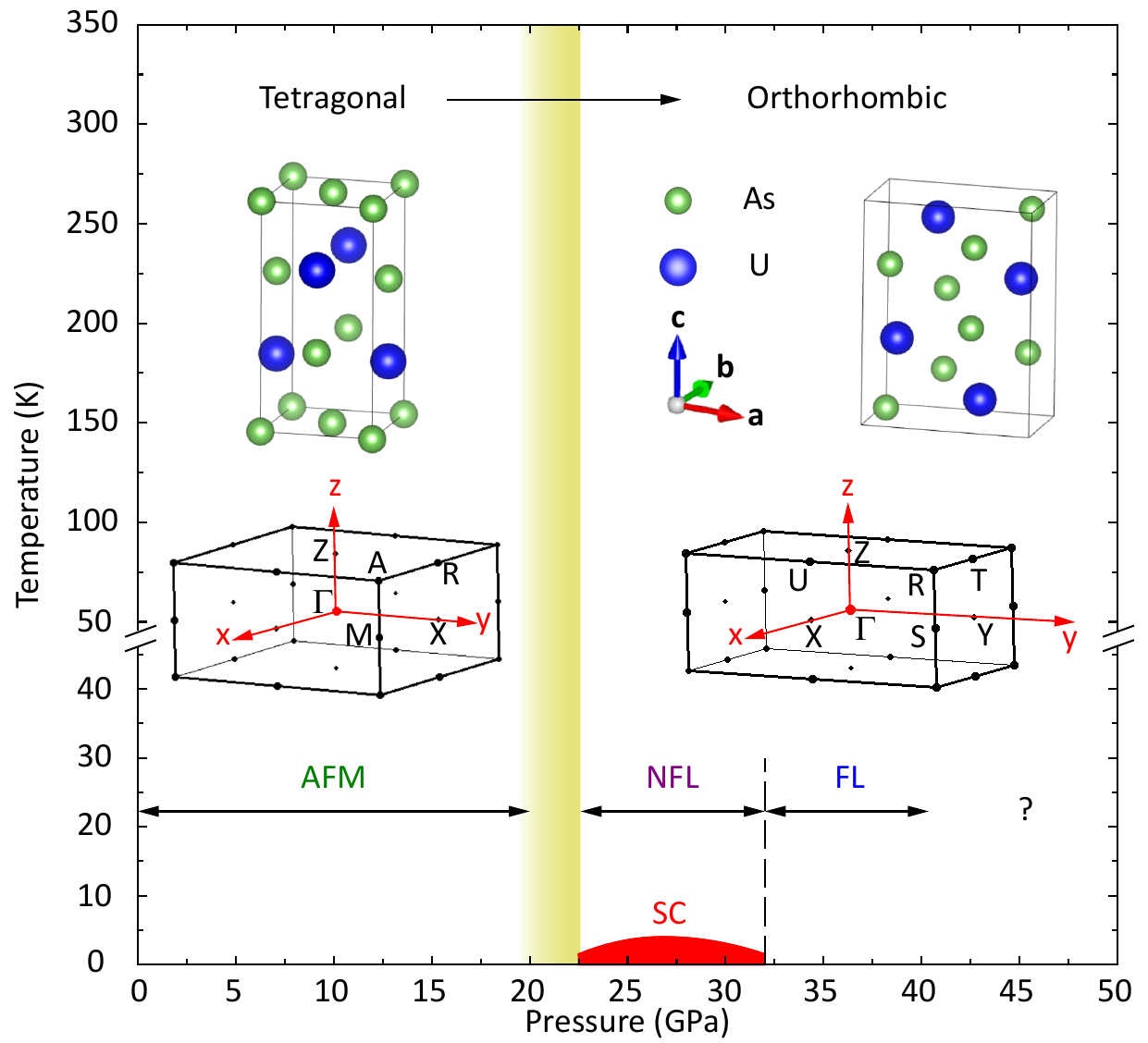}
\caption{Schematic phase diagram and crystal structures of $\mathrm{UAs_2}$ reproduced from Ref. \cite{Wen2026}. The upper panel shows the tetragonal and orthorhombic crystal structures together with their corresponding Brillouin zones in the low- and high-pressure phases, respectively. The lower panel presents the schematic pressure--temperature phase diagram below 40 K, showing the low-pressure antiferromagnetic (AFM) phase, the superconducting (SC) dome, the non-Fermi-liquid (NFL) region above $T_c$, and the high-pressure Fermi-liquid (FL) regime reported only up to 40.5 GPa. The yellow shaded region indicates the pressure range over which the tetragonal-to-orthorhombic structural transition occurs.}
		\label{fig1}
	\end{center}
\end{figure}

\begin{figure}[t]
	\begin{center}
		\includegraphics[width=0.48\textwidth]{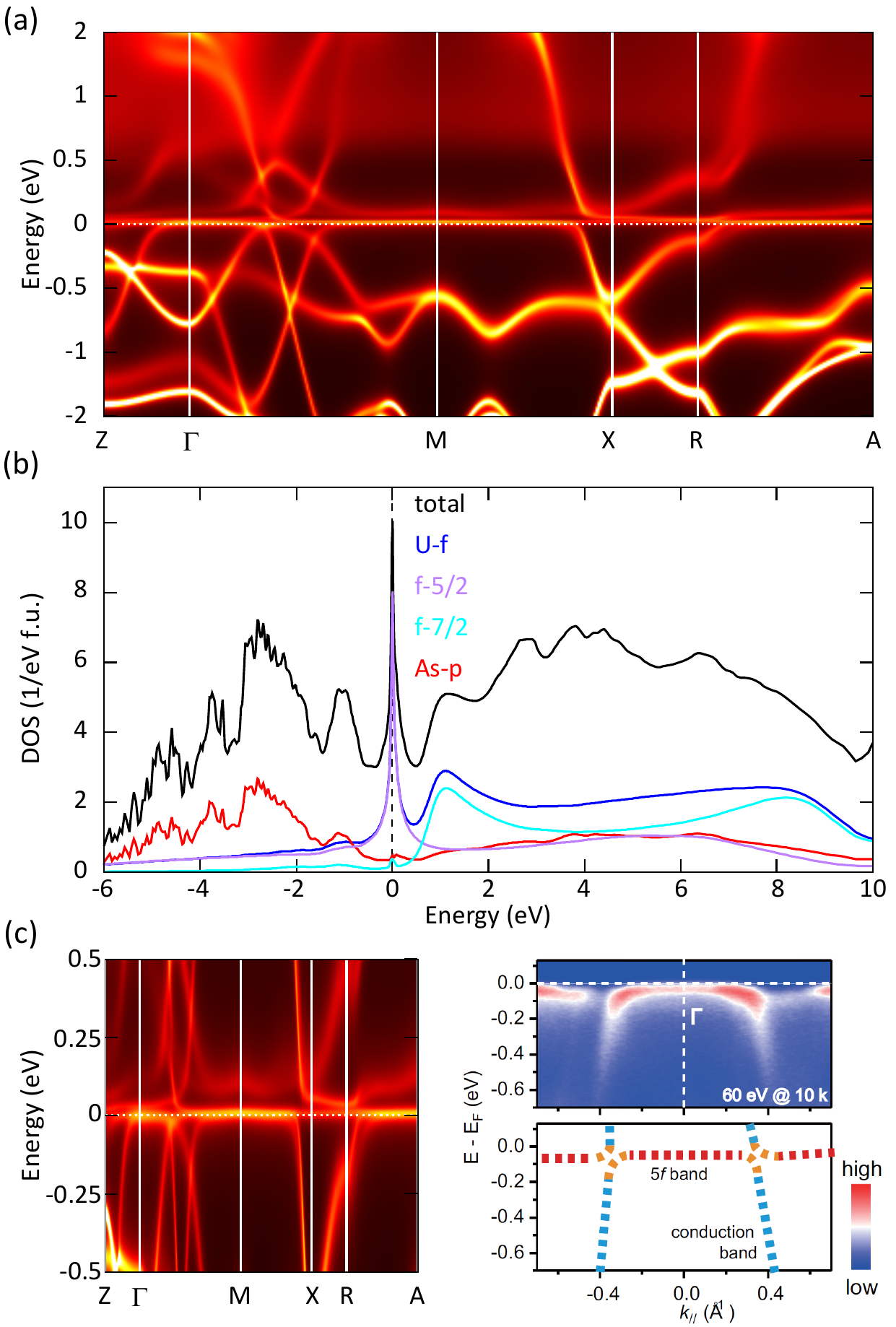}
\caption{Correlated electronic structure of $\mathrm{UAs_2}$ at ambient pressure obtained from DFT+DMFT at 50 K. (a) Momentum-resolved spectral function $A(\mathbf{k},\omega)$ along a high-symmetry path in the Brillouin zone. (b) Density of states showing the contributions from the U-$5f_{5/2}$, U-$5f_{7/2}$, and As-$3p$ orbitals. (c) Comparison between the calculated spectral function and ARPES measurements \cite{laiprb2022}, illustrating the origin of the flat bands along the $\Gamma$--$M$ direction.}
		\label{fig2}	
	\end{center}
\end{figure}

\textit{Struct.---} $\mathrm{UAs_2}$ crystallizes in a tetragonal anti-$\mathrm{Cu_2Sb}$-type structure with space group P4/nmm (No. 129, Z = 2) at ambient pressure, as shown in Fig.~1(a). The corresponding lattice parameters are $a = b = 3.96$ \AA\, and $c = 8.097$ \AA\,. Previous experimental studies have reported a structural transition at around $20-22$ GPa to a high-pressure orthorhombic phase with space group Pnma (No. 62, Z = 4), accompanied by a volume collapse of about 3$\%$ \cite{Wen2026}. The high-pressure crystal structure is shown in Fig.~1, with the lattice parameters $a = 6.4$ \AA\,, $b = 3.56$ \AA\,, and $c = 8.53$ \AA\, at 45 GPa \cite{Gerward1990}. The lattice constants at 26.8 and 30.8 GPa are obtained by interpolation based on the experimental volume evolution under pressure. The low- and high-pressure Brillouin zones and high-symmetry ${\bf k}$-points for band structure calculations are given in Fig.~\ref{fig1}.

\textit{Ambient pressure.---} We first examine the electronic structure of $\mathrm{UAs_2}$ at ambient pressure. Figure~\ref{fig2} summarizes the DFT+DMFT results obtained at $T=50$~K. The calculated spectral function in Fig.~\ref{fig2}(a) shows good agreement with previous ARPES measurements \cite{laiprb2022}. Notably, we see pronounced flat bands emerge in the vicinity of the $\Gamma$ and $M$ points, which arise from low-temperature Kondo hybridization between localized U-5$f$ electrons and conduction electrons, a hallmark of heavy-fermion physics. The partial densities of states (DOS) plotted in Fig.~\ref{fig2}(b) reveal that the flat bands near the Fermi energy are predominantly contributed by the U-5$f_{5/2}$ orbitals, while the 5$f_{7/2}$ orbitals exhibit broad peaks at higher energies. Figure~\ref{fig2}(c) further compares the calculated spectral function in a narrower energy window with the ARPES measurements. The DFT+DMFT results capture exactly the flat band features near the $\Gamma$ and $M$ points. We find multiple conduction bands participate in the hybridization along the $\Gamma$-$M$ direction, in contrast to previous speculation that the flat bands near $\Gamma$ and $M$ correspond to the upper and lower hybridization bands from a single conduction band, as illustrated by the dotted lines in the lower part of the right panel of Fig.~\ref{fig2}(c) reproduced from Ref. \cite{laiprb2022}. Our results support the multi-band scenario for $\mathrm{UAs_2}$.

\begin{figure}[t]
	\begin{center}
		\includegraphics[width=0.48\textwidth]{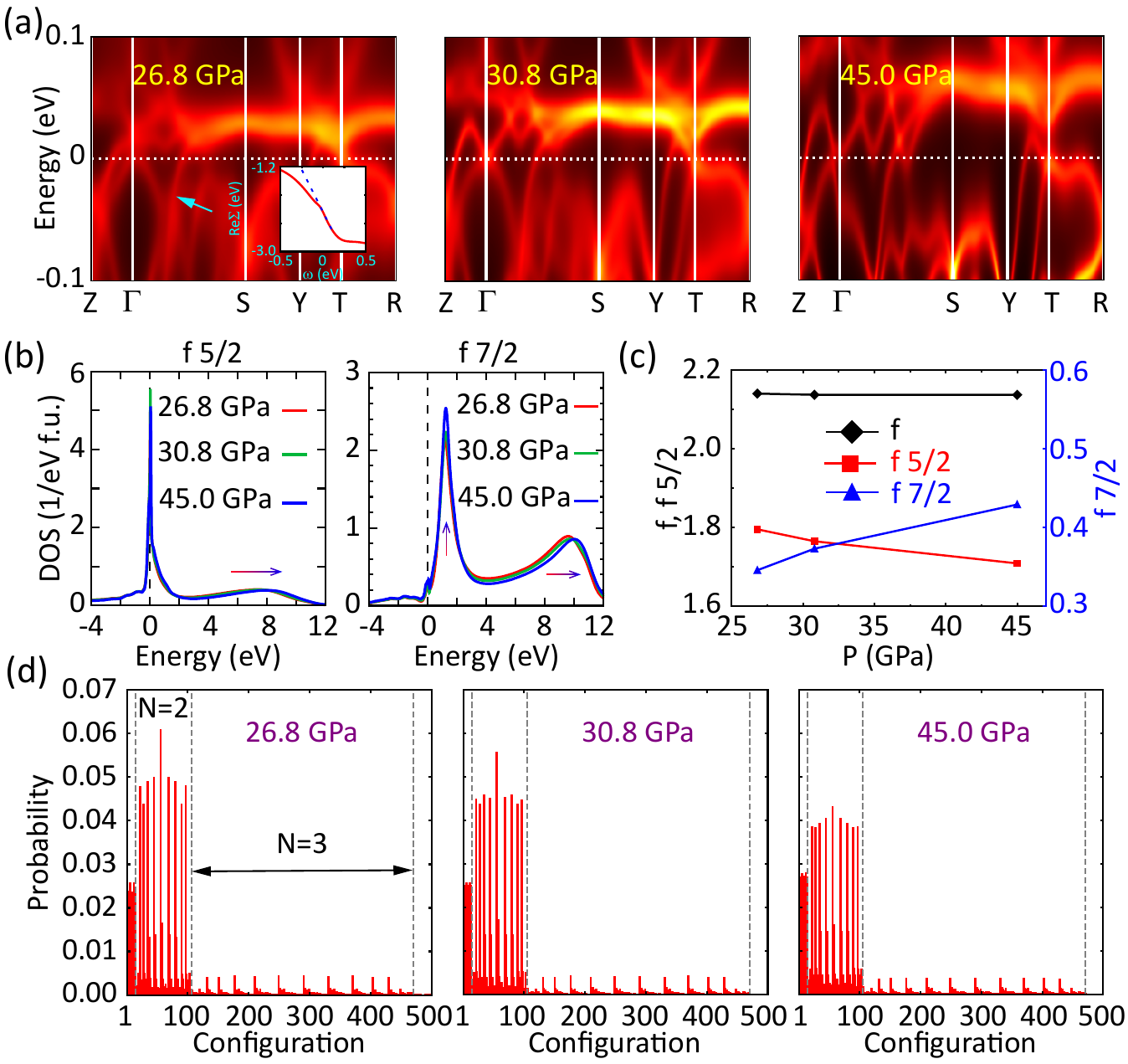}
\caption{Electronic structure of $\mathrm{UAs_2}$ under pressure obtained from DFT+DMFT at 50 K. (a) Momentum-resolved spectral functions at 26.8, 30.8, and 45 GPa. The arrow marks a low-energy kink, and the inset highlights the corresponding change in the slope of the real part of the real-frequency self-energy. (b) Pressure evolution of the U-$5f_{5/2}$ (left) and U-$5f_{7/2}$ (right) densities of states at 26.8, 30.8, and 45 GPa. (c) Pressure dependence of the U-$5f_{5/2}$, U-$5f_{7/2}$, and total U-$5f$ occupancies. (d) Probabilities of the atomic $5f$-electron configurations obtained from the CT-HYB impurity solver at 26.8, 30.8, and 45 GPa. The dashed lines separate the $N=1$, $2$, and $3$ configurations, where $N$ denotes the $5f$ electron number.}
		\label{fig3}
	\end{center}
\end{figure}

\begin{figure}[t]
	\begin{center}
		\includegraphics[width=0.5\textwidth]{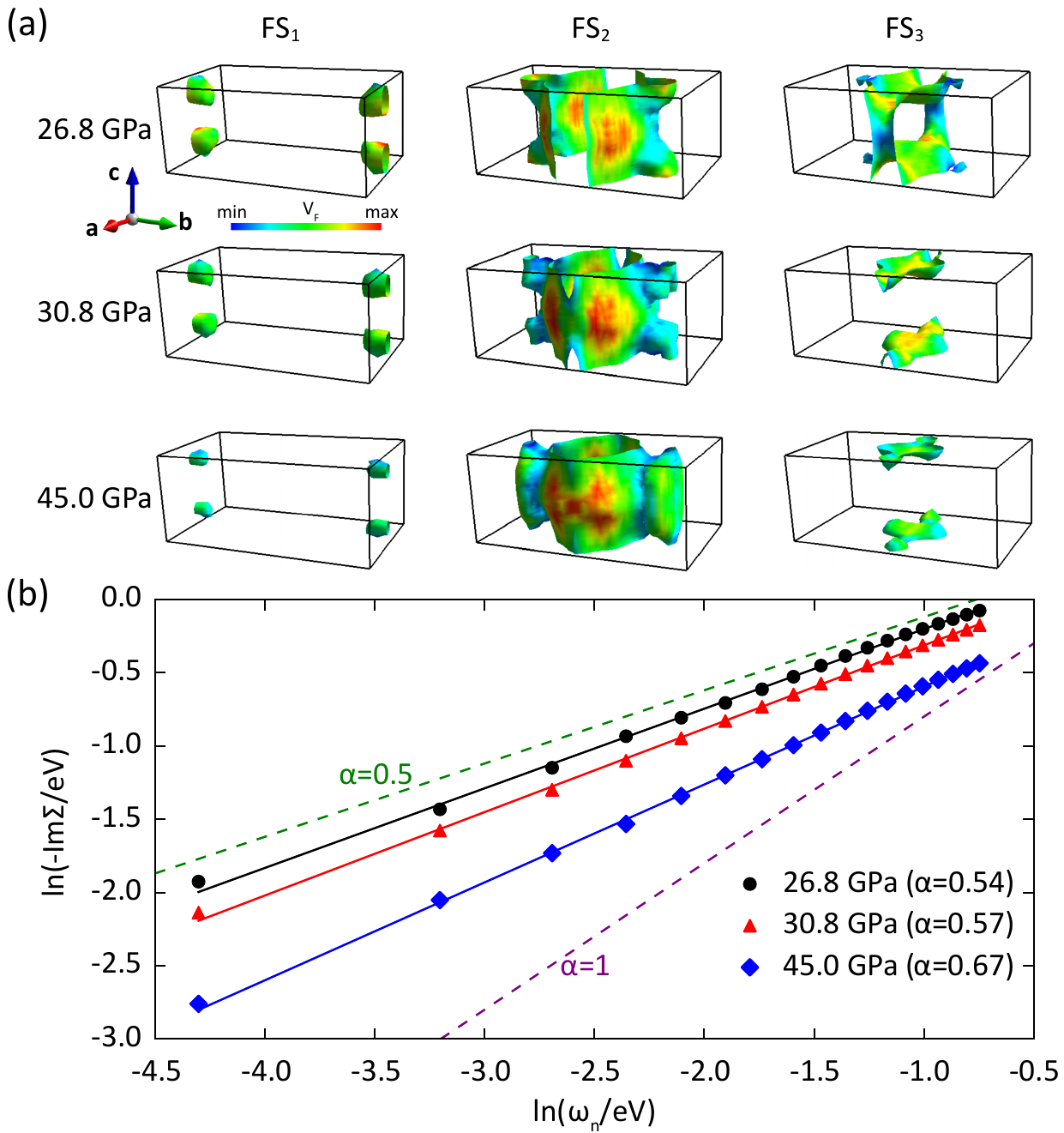}
\caption{(a) Evolution of the Fermi surfaces at 26.8, 30.8, and 45 GPa. (b) Log-log plots of the negative imaginary part of the Matsubara self-energy for the U-$5f_{5/2}$ states. The solid lines are power-law fits to the data at 26.8, 30.8, and 45 GPa, while the green and purple dashed lines denote the reference power laws with exponents of 0.5, characteristic of non-Fermi-liquid behavior, and 1, expected for a Fermi liquid, respectively. The Fermi-surface visualization was performed using the FermiSurfer code \cite{Kawamura2019CPC}. The colors represent the magnitudes of the Fermi velocity.}
		\label{fig4}
	\end{center}
\end{figure}

\textit{High Pressure.---} 
To investigate the evolution of the electronic structure under pressure, we further performed DFT+DMFT calculations for the high-pressure crystal structures at 26.8, 30.8, and 45 GPa at $T=50$ K. These pressures correspond to the maximum $T_{\mathrm{c}}$, a point near the upper edge of the superconducting dome, and a point well beyond the dome, respectively. As shown in Fig.~\ref{fig3}(a), the momentum-resolved spectral functions evolve markedly with increasing pressure. In contrast to ambient pressure, the flat hybridization bands gradually shift above the Fermi energy. In addition, as indicated by the arrow in Fig.~\ref{fig3}(a), a low-energy kink emerges in the spectral dispersion at about 40 meV below the Fermi energy, accompanied by a slope change in the real-frequency self-energy (inset). Such a low-energy kink is a genuine many-body feature previously predicted for Mott and Anderson lattice systems \cite{Byczuk2007NP,Hu2020PRR}. By contrast, over a wider energy window [Fig.~\ref{fig3}(b)], the overall density of states (DOS) remains similar to that at ambient pressure, except that the upper Hubbard band shifts slightly toward higher energy. The $5f_{5/2}$ DOS retains a sharp resonance near the Fermi energy, while the $5f_{7/2}$ DOS continues to exhibit two broad peaks above the Fermi energy. These results indicate that pressure predominantly modifies the low-energy electronic structure, while leaving the overall spectrum largely unchanged.

To uncover the microscopic origin of these low-energy changes, we next examine the evolution of the U-$5f$ electron occupancy [Fig.~\ref{fig3}(c)]. At 26.8 GPa, the occupations of the $f_{5/2}$ and $f_{7/2}$ manifolds are approximately 1.80 and 0.34, respectively, giving a total U-$5f$ occupancy of 2.14. As the pressure increases to 30.8 and 45 GPa, the $f_{5/2}$ occupation gradually decreases to 1.77 and 1.71, while the $f_{7/2}$ occupation increases correspondingly to 0.37 and 0.43, leaving the total U-$5f$ occupancy nearly unchanged. Thus, pressure does not change the total number of $5f$ electrons, but instead induces an electron transfer (self-doping) from the more localized $f_{5/2}$ to the more itinerant $f_{7/2}$ manifold, driving the $f_{5/2}$ electrons away from the Kondo regime toward a mixed-valence regime.

Figure~\ref{fig3}(d) compares the probabilities of different atomic configurations. Although the $N=2$ configurations ($N$ denotes the U-$5f$ electron number) remain dominant throughout the investigated pressure range, its probability decreases with increasing pressure, whereas those of the $N=1$ and $N=3$ configurations increase, indicating enhanced charge fluctuations. These results suggest that superconductivity in $\mathrm{UAs_2}$ is associated with pressure-induced self-doping and enhanced charge fluctuations, reminiscent of doping-controlled superconductivity in hole-doped cuprates \cite{MillisScience2000,KeimerNature2015,LeeRMP2006}, rather than the conventional Kondo lattice scenario in which  superconductivity is closely tied to magnetic quantum critical fluctuations of localized $f$-moments. A similar trend has previously been observed in $\mathrm{CeCu_2Si_2}$, where pressure enhances charge fluctuations and gives rise to a second superconducting dome with a maximum $T_c$ exceeding that at ambient pressure induced by antiferromagnetic spin fluctuations \cite{Yuan2003Science}. These similarities suggest that enhanced charge fluctuations may likewise contribute to the relatively high $T_c$ of pressurized $\mathrm{UAs_2}$ \cite{Li2021APS}.

Accompanying the self-doping and the upward shift of the flat hybridization bands, the Fermi surface undergoes a dramatic reconstruction. As illustrated in Fig.~\ref{fig4}(a), at 26.8 GPa, within the superconducting dome, the largest Fermi surface, i.e. FS$_2$, consists of two disconnected sheets with a quasi-two-dimensional geometry and enhanced nesting. As the pressure increases to 30.8 GPa, near the upper edge of the superconducting dome, these two sheets gradually bend toward each other and become connected. By 45 GPa, they have merged completely into a corrugated three-dimensional Fermi surface whose two-dimensional projection exhibits an electron-like topology reminiscent of heavily hole-doped cuprates \cite{DamascelliRMP2003,NormanNature1998}. These results suggest that the nested quasi-two-dimensional Fermi-surface topology may favor superconductivity in $\mathrm{UAs_2}$, whereas the suppression of superconductivity at higher pressures may be associated with a Lifshitz transition accompanying the Fermi-surface reconstruction \cite{Lifshitz1960}.

To further characterize the pressure evolution of the normal state, Fig.~\ref{fig4}(b) shows the log-log plots of the imaginary part of the Matsubara self-energy, $\ln[-\mathrm{Im}\Sigma(i\omega_n)]$ versus $\ln(\omega_n)$, for the U-5$f_{5/2}$ states at different pressures. The extracted slopes at 26.8, 30.8, and 45 GPa are 0.54, 0.57, and 0.67, respectively. These exponents lie between the reference values (dashed lines) of 0.5, characteristic of non-Fermi-liquid behavior, and 1, expected for a Fermi liquid. The systematic increase of the exponent with pressure indicates a gradual crossover from a non-Fermi-liquid state toward a Fermi-liquid-like state. Although it has not yet reached the Fermi liquid for unknown reason,  the overall trend is consistent with the pressure evolution of the experimental phase diagram shown in Fig.~\ref{fig1}. Thus, our DFT+DMFT calculations also provide a tentative picture for the pressure evolution of the normal state.

\textit{Conclusion.---}To summarize, we have investigated the electronic structure of $\mathrm{UAs_2}$ within the framework of DFT+DMFT and clarified its evolution under pressure. Our calculations show that the low-energy electronic states are dominated by strongly correlated U-$5f$ electrons, giving rise to flat hybridization bands near the Fermi energy that are consistent with spectroscopic observations at ambient pressure. Upon compression, the system undergoes a pressure-induced self-doping process, in which electrons are transferred from the more localized $j=5/2$ to the more itinerant $j=7/2$ manifold while the total U-$5f$ occupancy remains nearly unchanged. The self-doping results in enhanced charge fluctuations, a Fermi surface reconstruction, and a crossover from non-Fermi-liquid toward Fermi-liquid-like behavior. Remarkably, the evolution of these electronic characteristics closely follows the superconducting dome, suggesting that, in addition to potential magnetic fluctuations, enhanced charge fluctuations and favorable Fermi-surface topology may provide important microscopic ingredients for superconductivity in $\mathrm{UAs_2}$. Our work establishes a microscopic framework for understanding the interplay between electronic correlations, mixed valence, Fermi-surface topology, and superconductivity in uranium-based heavy-fermion systems.\\

This work was supported by the National Key Research and Development Program of China (Grants No. 2022YFA1402203 and No. 2024YFA1408602), the National Natural Science Foundation of China (Grant No. 12474136), and the High-Level Talent Research Start-Up Project Funding of Henan Academy of Sciences (Projects No. 20251827011 and No. 241827010). Numerical computations were performed at the Hefei Advanced Computing Center.


\begin{thebibliography}{99}
%introduction
\bibitem{StewartRMP1984}G. R. Stewart, Heavy-fermion systems, Rev. Mod. Phys. \textbf{56}, 755 (1984).
\bibitem{GegenwartNP2008}P. Gegenwart, Q. Si, and F. Steglich, Quantum criticality in heavy-fermion metals, Nat. Phys. \textbf{4}, 186 (2008).
\bibitem{MathurNat1998}N. D. Mathur, F. M. Grosche, S. R. Julian, I. R. Walker, D. M. Freye, R. K. W. Haselwimmer, and G. G. Lonzarich, Magnetically mediated superconductivity in heavy fermion compounds, Nature (London) \textbf{394}, 39 (1998).
\bibitem{PaschenNat2004}S. Paschen, T. Lühmann, S. Wirth, P. Gegenwart, O. Trovarelli, C. Geibel, F. Steglich, P. Coleman, and Q.Si, Hall-effect evolution across a heavy-fermion quantum critical point, Nature (London) \textbf{432}, 881 (2004).

\bibitem{SaxenaNat2000}S. S. Saxena, P. Agarwal, K. Ahilan, F. M. Grosche, R. K. W. Haselwimmer, M. J. Steiner, E. Pugh, I. R. Walker, S. R. Julian, P. Monthoux, G. G. Lonzarich, A. Huxley, I. Sheikin, D. Braithwaite, and J. Flouquet, Superconductivity on the border of itinerant-electron ferromagnetism in UGe$_2$, Nature (London) 406, 587 (2000).
\bibitem{JoyntRMP2002}R. Joynt and L. Taillefer, The superconducting phases of UPt$_3$, Rev. Mod. Phys. \textbf{74}, 235 (2002).

%UTe2
\bibitem{RanSci2019}S. Ran, C. Eckberg, Q.-P. Ding, Y. Furukawa, T. Metz, S. R. Saha, I.-L. Liu, M. Zic, H. Kim, J. Paglione, and N. P. Butch, Nearly ferromagnetic spin-triplet superconductivity, Science \textbf{365}, 684 (2019).
\bibitem{AokiJpsj2019}D. Aoki, A. Nakamura, F. Honda, D. Li, Y. Homma, Y. Shimizu, Y. J. Sato, G. Knebel, J.-P. Brison, A. Pourret, D. Braithwaite, G. Lapertot, Q. Niu, M. Vali\v{s}ka, H. Harima, and J. Flouquet, Unconventional superconductivity in heavy fermion UTe$_2$, J. Phys. Soc. Jpn. \textbf{88}, 043702 (2019).
\bibitem{XuPRL2019}Y. Xu, Y. Sheng, and Y.-F. Yang, Quasi-two-dimensional Fermi surfaces and unitary spin-triplet pairing in the heavy fermion superconductor UTe$_2$, Phys. Rev. Lett. \textbf{123}, 217002 (2019).

%UAu2
\bibitem{O’NeillPnas2022}C. D. O’Neill, J. L. Schmehr, and A. D. Huxley, Multicomponent odd-parity superconductivity in UAu$_2$ at high pressure, Proc. Natl. Acad. Sci. USA \textbf{119}, e2210235119 (2022).

%UX2
\bibitem{AmorettiJmmm1984}G. Amoretti, A. Blaise, and J. Mulak, Crystal field interpretation of the magnetic properties of UX$_2$ compounds (X = P, As, Sb, Bi), J. Magn. Magn. Mater. \textbf{42}, 65 (1984).
\bibitem{Gerward1990}L. Gerward, J. S. Olsen, U. Benedict, S. Dabos-Seignon, and H. Luo, Crystal structures of UP$_2$, UAs$_2$, UAsS, and UAsSe in the pressure range up to 60 GPa. High Temp. High Press. \textbf{22}, 523 (1990).


\bibitem{ChenPRL2019}Q. Y. Chen, X. B. Luo, D. H. Xie, M. L. Li, X. Y. Ji, R. Zhou, Y. B. Huang, W. Zhang, W. Feng, and Y. Zhang, Orbitalselective Kondo entanglement and antiferromagnetic order in USb$_2$, Phys. Rev. Lett. \textbf{123}, 106402 (2019).
\bibitem{GiannakisSciAdv2019}I. Giannakis, J. Leshen, M. Kavai, S. Ran, C. J. Kang, S. R. Saha, Y. Zhao, Z. Xu, J. W. Lynn, L. Miao, L. A. Wray, G. Kotliar, N. P. Butch, and P. Aynajian, Orbital-selective Kondo lattice and enigmatic f electrons emerging from inside the antiferromagnetic phase of a heavy fermion, Sci. Adv. \textbf{5}, eaaw9061 (2019).

\bibitem{FengPRB2021}W. Feng, D. H. Xie, X. B. Luo, S. Y. Tan, Y. Liu, Q. Liu, Q. Q. Hao, X. G. Zhu, Q. Zhang, Y. Zhang, Q. Y. Chen, and X. C. Lai, Crossover behavior of the localized to itinerant transition of 5 f electrons in the antiferromagnetic Kondo lattice USb$_2$, Phys. Rev. B \textbf{104}, 235103 (2021).
\bibitem{MiaoNc2019}L. Miao, R. Basak, S. Ran, Y. S. Xu, E. Kotta, H. W. He, J. D. Denlinger, Y. D. Chuang, Y. Zhao, Z. Xu, J. W. Lynn, R. Jeffries, S. R. Saha, I. Giannakis, P. Aynajian, C. J. Kang, Y. L. Wang, G. Kotliar, N. P. Butch, and L. A. Wray, High temperature singlet-based magnetism from Hund’s rule correlations, Nat. Commun. \textbf{10}, 8 (2019).
\bibitem{SiddiqueeNc2023}H. Siddiquee, C. Broyles, E. Kotta, S. Liu, S. Peng, T. Kong, B. Kang, Q. Zhu, Y. Lee, L. Ke, H. Weng, J. D. Denlinger, L. A. Wray, and S. Ran, Breakdown of the scaling relation of anomalous Hall effect in Kondo lattice ferromagnet USbTe, Nat. Commun. \textbf{14}, 527 (2023).


%UAs2
\bibitem{laiprb2022}X. Ji, X. Luo, Q. Chen, W. Feng, Q. Hao, Q. Liu, Y. Zhang, Y. Liu, X. Wang, S. Tan, and X. Lai, Direct observation of coexisting Kondo hybridization and antiferromagnetic state in UAs$_2$, Phys. Rev. B \textbf{106}, 125120 (2022). 
\bibitem{JiPRB2024}X. Ji, Q. Liu, W. Feng, Y. Zhang, Q. Chen, Y. Liu, Q. Hao, J. Wu, Z. Xue, X. Zhu, Q. Zhang, X. Luo, S. Tan, and X. Lai, Heavy fermion related behaviors and the effects from nonmagnetic atom vacancies in the 5f-electron based antiferromagnet UAs$_2$, Phys. Rev. B \textbf{109}, 075158 (2024).

%UAs2 SuperCon
\bibitem{Wen2026}Q. Li, Z. -N. Xiang, B. -B Zhang, Y. -J. Zhang, C. Zhang, and H. -H. Wen, Unconventional superconductivity in UAs$_2$ under pressure, Sci. Adv. \textbf{12 (15)}, eaed6248(2026)

%DFT
\bibitem{2014WIEN2k} P. Blaha, K. Schwarz, G. K. H. Madsen, D. Kvasnicka, and J. Luitz, WIEN2k, an augmented plane wave + local orbitals program for calculating crystal properties (Karlheinz Schwarz, Techn. Universit{\"a}t Wien, Austria), (2001). ISBN 3-9501031-1-2.

\bibitem{WIEN2k} P. Blaha, K. Schwarz, F. Tran, R. Laskowski, G. K. H. Madsen, and L. D. Marks, WIEN2k: An APW+lo program for calculating the properties of solids, J. Chem. Phys. \textbf{152}, 074101 (2020).



%DMFT
\bibitem{Georges1996RMP} A. Georges, G. Kotliar, W. Krauth, and M. J. Rozenberg, Dynamical mean-field theory of strongly correlated fermion systems and the limit of infinite dimensions, Rev. Mod. Phys. \textbf{68}, 13 (1996).

\bibitem{Anisimov1997JPCM} V. I. Anisimov, A. I. Poteryaev, M. A. Korotin, A. O. Anokhin, and G. Kotliar, First-Principles Calculations of the Electronic Structure and Spectra of Strongly Correlated Systems: Dynamical Mean-Field Theory, J. Phys.: Condens. Matter \textbf{9}, 7359 (1997).

\bibitem{Lichtenstein1998PRB} A. I. Lichtenstein and M. I. Katsnelson, Ab initio calculations of quasiparticle band structure in correlated systems: LDA++ approach, Phys. Rev. B \textbf{57}, 6884 (1998).

\bibitem{Kotliar2006RMP} G. Kotliar, S. Y. Savrasov, K. Haule, V. S. Oudovenko, O. Parcollet, and C. A. Marianetti, Electronic structure calculations with dynamical mean-field theory, Rev. Mod. Phys. \textbf{78}, 865 (2006).

\bibitem{Held2008JPCM} K. Held, O. K. Andersen, M. Feldbacher, A. Yamasaki, and Y.-F. Yang, Band structure meets many-body theory: the LDA+DMFT method, J. Phys.: Condens. Matter \textbf{20}, 064202 (2008).

\bibitem{Haule2010PRB} K. Haule, C.-H. Yee, and K. Kim, Dynamical mean-field theory within the full-potential methods: Electronic structure of {CeIrIn$_5$}, {CeCoIn$_5$}, and {CeRhIn$_5$}, Phys. Rev. B {\bf 81}, 195107 (2010).

%GGA
\bibitem{Perdew1996PRL} J. P. Perdew, K. Burke, and M. Ernzerhof, Generalized gradient approximation made simple, Phys. Rev. Lett. \textbf{77}, 3865 (1996).

%CTQMC solver
\bibitem{Haule2007PRB} K. Haule, Quantum Monte Carlo impurity solver for cluster dynamical mean-field theory and electronic structure calculations with adjustable cluster base, Phys. Rev. B \textbf{75}, 155113 (2007).

%UJ for UX
\bibitem{ShimEl2009} J. H. Shim, K. Haule, and G. Kotliar, X-ray absorption branching ratio in actinides: LDA+DMFT approach, Europhys. Lett. \textbf{85}, 17007 (2009).
\bibitem{YinPRB2011} Q. Yin, A. Kutepov, K. Haule, G. Kotliar, S. Y. Savrasov, and W. E. Pickett, Electronic correlation and transport properties of nuclear fuel materials, Phys. Rev. B \textbf{84}, 195111 (2011).

%DMFT doublecounting
\bibitem{Haule2015PRL} K. Haule, Exact double counting in combining the dynamical mean field theory and the density functional theory, Phys. Rev. Lett. \textbf{115}, 196403 (2015).

%the maximum entropy
\bibitem{Jarrell1996PR} M. Jarrell and J. E. Gubernatis, Bayesian inference and the analytic continuation of imaginary-time quantum Monte Carlo data, Phys. Rep. \textbf{269}, 133 (1996).

%kink
\bibitem{Byczuk2007NP}K. Byczuk, M. Kollar, K. Held, Yi-feng Yang, I. A. Nekrasov, Th. Pruschke, and D. Vollhardt, Kinks in the dispersion of strongly correlated electrons. Nature Physics \textbf{3}, 168 (2007).

\bibitem{Hu2020PRR}D. Hu, N.-H. Tong, and Y.-F. Yang, Energy-scale cascade and correspondence between Mott and Kondo lattice physics. Phys. Rev. Research \textbf{2}, 043407 (2020).



%self doping
\bibitem{MillisScience2000}J. Orenstein and A. J. Millis, Advances in the physics of high-temperature superconductivity, Science \textbf{288}, 468 (2000).
\bibitem{KeimerNature2015}B. Keimer, S. A. Kivelson, M. R. Norman, S. Uchida, and J. Zaanen, From quantum matter to high-temperature superconductivity in copper oxides, Nature (London) \textbf{518}, 179 (2015).
\bibitem{LeeRMP2006}P. A. Lee, N. Nagaosa, and X.G. Wen, Doping a Mott insulator: Physics of high-temperature superconductivity, Rev. Mod. Phys. \textbf{78}, 17 (2006).

%heavy fermion
\bibitem{Yuan2003Science}H. Q. Yuan, F. M. Grosche, M. Deppe, C. Geibel, G. Sparn, and F. Steglich, Observation of Two Distinct Superconducting Phases in CeCu$_2$Si$_2$. Science \textbf{302}, 2104-2107 (2003).
\bibitem{Li2021APS}Y. Li, Y.-T. Sheng, and Y.-F. Yang, Theoretical progress and material studies of heavy fermion superconductors. Act. Phys. Sin. \textbf{70}, 071402 (2021).

%fermisurfer
\bibitem{Kawamura2019CPC} M. Kawamura, FermiSurfer: Fermi-surface viewer providing multiple representation schemes. Comp. Phys. Commun \textbf{239}, 197 (2019).


%FermiSurace
\bibitem{DamascelliRMP2003}A. Damascelli, Z. Hussain, and Z.-X. Shen, Angle-resolved photoemission studies of the cuprate superconductors, Rev. Mod. Phys. \textbf{75}, 473 (2003).
\bibitem{NormanNature1998}M. R. Norman, H. Ding, M. Randeria, J.C. Campuzano, T. Yokoya, T. Takeuchi, T. Takahashi, T. Mochiku, K. Kadowaki, P. Guptasarma, and D.G. Hinks, Destruction of the Fermi surface in underdoped high-$T_c$ superconductors, Nature (London) \textbf{392}, 157 (1998).

%LifshitzTransition
\bibitem{Lifshitz1960}I. M. Lifshitz, Anomalies of electron characteristics of a metal in the high pressure region, Sov. Phys. JETP \textbf{11}, 1130 (1960).




\end{thebibliography}
\end{document}